\documentclass[apj,numberedappendix]{emulateapj}

\def\gtorder{\mathrel{\raise.3ex\hbox{$>$}\mkern-14mu
             \lower0.6ex\hbox{$\sim$}}}
\def\ltorder{\mathrel{\raise.3ex\hbox{$<$}\mkern-14mu
             \lower0.6ex\hbox{$\sim$}}}

\slugcomment{Accepted to ApJ}

\shorttitle{Radio sources variability}

\shortauthors{Ofek \& Frail}

\begin{document}

\title{The structure function of variable 1.4\,GHz radio sources based on NVSS and FIRST observations}
\author{Eran~O.~Ofek\altaffilmark{1}$^{,}$\altaffilmark{2},
and Dale~A.~Frail\altaffilmark{3}}

\altaffiltext{1}{Division of Physics, Mathematics and Astronomy, California Institute of Technology, Pasadena, CA 91125, USA}
\altaffiltext{2}{Einstein fellow}
\altaffiltext{3}{National Radio Astronomy Observatory, P.O. Box O, Socorro, NM 87801, USA}

\begin{abstract}


We augment the two widest/deepest 1.4\,GHz radio surveys: the NRAO VLA Sky
Survey (NVSS) and the Faint Images of the Radio Sky at Twenty-Centimeters
(FIRST), with the mean epoch in which each source was observed. We use these
catalogs to search for unresolved sources which vary between the FIRST and NVSS
epochs. We find 43 variable sources (0.1\% of the sources) which vary by more
than 4$\sigma$, and we construct the mean structure function of these objects.
This enables us to explore radio variability on time scales between several
months and about five years. We find that on these time scales, the mean
structure function of the variable sources is consistent with a flat
structure function.
A plausible explanation to these observations is that a large fraction of the
variability at 1.4\,GHz is induced by scintillations in the interstellar
medium, rather than by intrinsic variability. Finally, for a sub sample
of the variables for which the redshift is available, we do~not find strong
evidence for a correlation between the variability amplitude and the source
redshift.

\end{abstract}


\section{Introduction}
\label{Intro}

Variability of radio sources at low frequencies is mainly attributed
to propagation effects (scintillations) induced by large scale
electron density inhomogeneities in the Inter-Stellar Medium (ISM;
e.g., Hunstead 1972; 
Rickett, Coles \& Bourgois 1984;
Rickett 1990; Ghosh \& Rao 1992).
The predicted variability structure function of compact radio sources
due to scintillations (e.g., Blandford \& Narayan 1985; Goodman \&
Narayan 1985; Blandford et al. 1986; Hjellming \& Narayan 1986) is
roughly consistent with the typically observed structure function, at
least below 5\,GHz (e.g., Qian et al. 1995; Gaensler \& Hunstead
2000).  These models predict a rise in the structure function
up to time scales of $\sim 10$\,days at $\approx 5$\,GHz and up to
$\sim 100$\,days at $\approx 0.5$\,GHz, followed by a flattening of
the structure function.

Specifically, Qian et al. (1995) analyzed radio observations of the
compact radio source 1741$-$038 ($z=1.054$) taken in several
frequencies between 1.5 and 22\,GHz.  They compared the observed
structure functions with theoretical models
for scattering by an extended Galactic medium, with and without
a thin screen component.
They reported that for frequencies below about
5\,GHz the observations are consistent with a scattering by an
extended Galactic medium and a thin screen.  However, above this
frequency they found excess variability relative to the models.
Moreover, at these high frequencies, the structure function continues
to rise towards longer time scales.  They suggested that at high
frequencies ($\gtorder 5$\,GHz) some of the variability of this radio
source is intrinsic.
This general picture is also supported by Mitchell et al. (1994).

Gaensler \& Hunstead (2000) studied the variability of 55 radio
calibrators observed by the Molonglo Observatory Synthesis Telescope
(MOST) at 843\,MHz.  They constructed the structure function for 18
variable sources.  For the majority of these variable objects the
structure function flattens on time scales of a few hundreds days.
Furthermore, they confirmed early results (Condon et al. 1979) which
found that the variability amplitude is increasing as a function of
the source spectral index $\alpha$, defined by $f_{\nu}\propto
\nu^{\alpha}$, where $f_{\nu}$ is the specific flux at frequency
$\nu$.  This is attributed to a correlation between the source angular
size and spectral index.
Another confirmation for the importance of Galactic scintillations is
that radio variability
depends on Galactic latitude (e.g., Spangler et al. 1989; Ghosh \& Rao
1992; Gaensler \& Hunstead 2000).  We note however that Rys \&
Machalski (1990) did~not find evidence for increasing fraction of
1.4\,GHz variability for sources brighter than 100\,mJy at low
Galactic latitudes.

Lovell et al. (2008) presented results from the Micro-Arcsecond
Scintillation-Induced Variability (MASIV) survey conducted at 5\,GHz.
Among their findings: half of the sources they monitored exhibit
2\%--10\% rms variations on time scales over two days. They also found
that the structure function of the variable sources rises on time
scales of a few days, and the variability amplitudes correlates with
the H$\alpha$ emission at the direction of the sources. Furthermore,
there is evidence that the variability amplitude decrease with
redshift above $z\approx2$, presumably due to evolution of the source
size with redshift (see however Lazio et al.~2008).

Here we compare the two widest/deepest 1.4\,GHz sky surveys taken using
the Very Large Array\footnote{The Very
  Large Array is operated by the National Radio Astronomy Observatory (NRAO),
  a facility of the National Science Foundation operated under
  cooperative agreement by Associated Universities, Inc.} (VLA),
and search for variable sources.
We use these datasets to construct the average
structure function on time scales between several months
and about five years
and to look for indications for intrinsic variability of these
sources.
In \S\ref{Cat} we present augmented versions of the FIRST and NVSS
catalogs which contains the mean time in which each
source was observed.
In \S\ref{CrossCorr} we cross correlate the two catalogs,
and in \S\ref{SF} we construct the structure function
of the variable sources.
Finally, we discuss the results in \S\ref{Disc}.

\section{The catalogs}
\label{Cat}

The NVSS observations were carried out between June 1993
and April 1999, while the FIRST survey observations were conducted
between March 1993 and September 2002.  Therefore, these observations
provide a long baseline to the structure function analysis.
Constructing the structure function of variable sources requires
knowledge of their fluxes at multiple epochs and the time at which the
observations were taken.  However, the NRAO VLA Sky Survey (NVSS;
Condon et al. 1998) and the Faint Images of the Radio Sky at
Twenty-Centimeters (FIRST; Becker et al. 1995) source catalogs do~not contain
the time at which each source was observed.  The reason for this is
that the observing times are not well defined.  Images in both surveys
were taken by scanning the sky in a hexagonal grid in which observing
points are separated by $26'$.  Both surveys were obtained using the
VLA, in which the full width at half power at 1.4\,GHz is $\cong31'$.
Each primary beam field of view was truncated to
radii of $24'$ and $23.5'$ for the
NVSS and FIRST surveys, respectively.
Therefore, each point on the sky
effectively contains information from roughly four different pointings
taken at different times.

In order to obtain the epoch at which each source was observed we
downloaded from the VLA
archive\footnote{https://archive.nrao.edu/archive/advquery.jsp} the
list of all observing scans which are associated with each sky
survey\footnote{These are observing code AC308 for the NVSS catalog
  and AB628, AB879 and AB950 for the FIRST catalog.}.

Next, we cross-correlate the list of observing scans for each project
with its catalog of sources.  We use a matching radii equal to the
truncation radii of $24'$ and $23.5'$ for the NVSS and
FIRST\footnote{We use FIRST catalog version 20080716.} surveys,
respectively.  This enable us to estimate for each source, the number
of observations ($N_{{\rm obs}}$), its mean observing time (over all
scan mid times, $t_{{\rm ep}}$) and the time span within which these
observation were obtained ($\delta{t}$).

The products are versions of the FIRST and NVSS catalogs that contains
the observing time, number of observations, and time span of
observations for each source\footnote{These are approximate
observing times since we do~not know if all the data was
used in the reduction process of the FIRST and NVSS.}.
In Tables~\ref{Tab:NVSStime}--\ref{Tab:FIRSTtime} we present a version of
these catalogs containing the source coordinates and the observing
time information for each source.
\begin{deluxetable}{lllllll}
\tablecolumns{7}
\tabletypesize{\scriptsize}
\tablewidth{0pt}
\tablecaption{Observing times of NVSS sources}
\tablehead{
\colhead{J2000 RA}   &
\colhead{J2000 Dec}  &
\colhead{$f$}  &
\colhead{$\sigma_{{\rm f}}$}  &
\colhead{$N_{{\rm obs}}$} &
\colhead{$t_{{\rm ep}}$} &
\colhead{$\delta{t}$} \\
\colhead{deg}   &
\colhead{deg}   &  
\colhead{mJy}   &  
\colhead{mJy}   &  
\colhead{}      &
\colhead{days}  &
\colhead{days}     
}
\startdata
  194.89704 & $-40.37908$ &       3.60 &      0.70 &     1&  0220.698 &  0.000\\
  249.00512 & $-40.37366$ &      99.70 &      3.60 &     1&  0220.840 &  0.000\\
  199.42008 & $-40.36441$ &      24.30 &      1.20 &     1&  0220.699 &  0.000\\
  248.90470 & $-40.36169$ &       3.00 &      0.60 &     1&  0220.840 &  0.000\\
  212.86787 & $-40.35236$ &       2.70 &      0.50 &     1&  0220.737 &  0.000
\enddata
\tablecomments{A version of the NVSS catalog containing the source position,
mean observing time ($t_{{\rm ep}}$),
number of scans ($N_{{\rm obs}}$),
and the time span over which the scans were obtained ($\delta{t}$).
The observing time, $t_{{\rm ep}}$, is given in
${\rm JD}-2450000$ days,
where JD is the Julian day.
$f$ is the peak flux density and $\sigma_{{\rm f}}$ is
the error in the peak flux density.
We note that our cross correlation of the observing scans and the sources
catalog was not able to produce the observing times of 591 sources.
This table is published in its entirety in the electronic edition of
the {\it Astrophysical Journal}. A portion of the full table is shown here for
guidance regarding its form and content.
The table is sorted by Declination, therefore the sources
listed here are near the edge of the survey footprint and have a single
observation.}
\label{Tab:NVSStime}
\end{deluxetable}
%
\begin{deluxetable}{lllllll}
\tablecolumns{7}
\tabletypesize{\scriptsize}
\tablewidth{0pt}
\tablecaption{Observing times of FIRST sources}
\tablehead{
\colhead{J2000 RA}   &
\colhead{J2000 Dec}  &
\colhead{$f$}  &
\colhead{$\sigma_{{\rm f}}$}  &
\colhead{$N_{{\rm obs}}$} &
\colhead{$t_{{\rm ep}}$} &
\colhead{$\delta{t}$} \\
\colhead{deg}   &
\colhead{deg}   &  
\colhead{mJy}   &  
\colhead{mJy}   &  
\colhead{}      &
\colhead{days}  &
\colhead{days}     
}
\startdata
  354.74904 &  $-11.39882$&    1.68  &     0.14 &     1&  0593.087 &  0.000\\
    5.19669 &  $-11.39761$&    1.02  &     0.15 &     1&  0593.118 &  0.000\\
    6.82096 &  $-11.39611$&    1.11  &     0.15 &     1&  0595.073 &  0.000\\
  359.90417 &  $-11.39585$&    1.39  &     0.15 &     1&  0593.102 &  0.000\\
    6.70133 &  $-11.39458$&    1.17  &     0.14 &     1&  0595.073 &  0.000
\enddata
\tablecomments{Like Table~\ref{Tab:NVSStime} but for the FIRST catalog.
We note that our cross correlation of the observing scans and the sources
catalog was not able to produce the observing times of 379 sources.}
\label{Tab:FIRSTtime}
\end{deluxetable}

\begin{figure}
\centerline{\includegraphics[width=8.5cm]{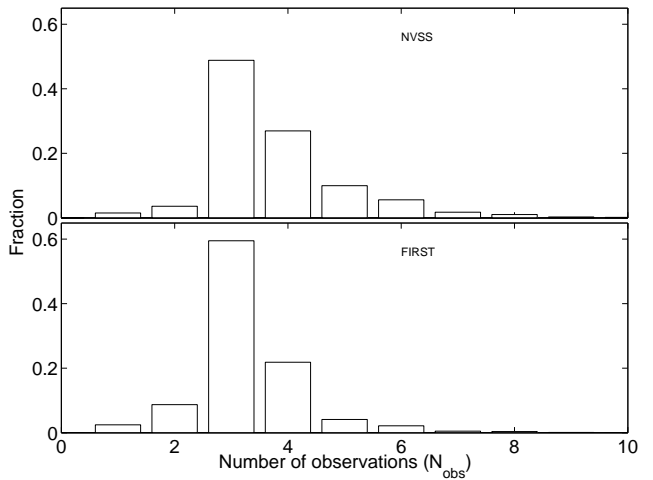}}
\caption{Histogram of number
of observations from which each source image
was composed.
Bin size is 1 observation.
The upper panel is for NVSS sources, while
the lower panel is for FIRST sources.
\label{Fig:Hist_Nobs_FIRST_NVSS}}
\end{figure}

\begin{figure}
\centerline{\includegraphics[width=8.5cm]{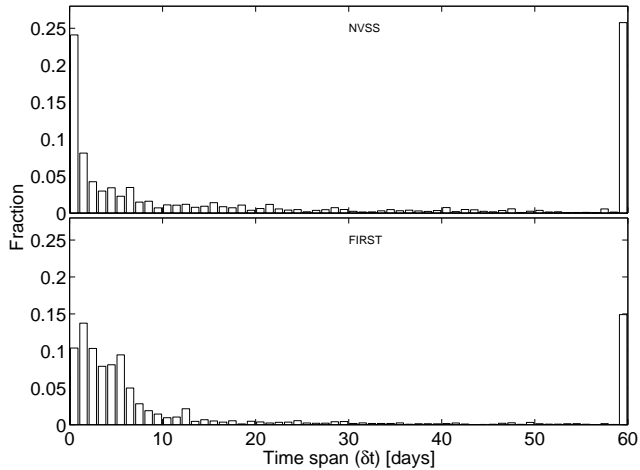}}
\caption{Histogram of
the time span of observations from which each source image
was composed ($\delta{t}$).
Bin size is 1~day.
Panels like in Fig.~\ref{Fig:Hist_Nobs_FIRST_NVSS}.
The last bin on the right-hand side is for all $\delta{t}>59$\,days.
\label{Fig:Hist_Range_FIRST_NVSS}}
\end{figure}

Figure~\ref{Fig:Hist_Nobs_FIRST_NVSS} presents histograms of the
number of individual snapshots used to compose each source image.
For both surveys, the typical number of snapshots per composite image
is 3--4 (see also Helfand et al. 1996).
In Figure~\ref{Fig:Hist_Range_FIRST_NVSS} we show histograms of
$\delta{t}$ for the two surveys.  This figure suggests that most of
the images are made from snapshots taken within a few weeks range of
each other.

We note that a transient search based on the comparison of the FIRST
and NVSS catalogs was presented in Levinson et al. (2002) and
discussed in Gal-Yam et al. (2006) and Ofek et al. (2010).  However,
these previous efforts did~not use the observing time of the
sources.

\section{Cross correlation of the FIRST and NVSS catalogs}
\label{CrossCorr}

The FIRST catalog contains 816,331 sources,
brighter than about $1$\,mJy, mainly in the North
Galactic cap.
About $81$\% of these sources have $\delta{t}<30$\,days.
The NVSS catalog contains 1,773,484 objects
with $\delta>-30$\,deg,
brighter than about $3.5$\,mJy, of which
$\cong 67$\% have $\delta{t}<30$\,days.
We select all the FIRST sources which deconvolved
major and minor axes equal zero (i.e., point sources),
the peak flux density is
larger than 5\,mJy, all the scans composing
their flux measurement ($\delta{t}$) were taken within 30 days,
and which are isolated from any other FIRST source (of any kind)
within $60''$.
Since the resolution of the NVSS is $45''$, nine times
coarser than that of the FIRST survey,
the last step is designed to remove
NVSS sources which flux may be contaminated by multiple FIRST objects.
We found 6463 FIRST sources that satisfy these criteria.
Next, from the NVSS catalog we select all sources 
with $\delta{t}<30$\,days -- we find
1,183,620 sources that satisfy this criterion.

Then, for each object in the subset of the FIRST catalog we search
for a source in the subset of the NVSS catalog which is found within
$15''$ of the FIRST object\footnote{The median astrometric error for
  5\,mJy sources in the NVSS catalog is about $3''$, and 99.4\% of the
  errors of such sources are smaller than $15''$}.  We found 4367
matched sources.  These matches represent point sources for which we
have both a FIRST and NVSS flux-density measurements.
We note that only $68\%$ of the FIRST sources in this list have NVSS
matches. This is mostly due to the fact that we used only NVSS
sources with $\delta{t}<30$\,days.

\begin{figure}
\centerline{\includegraphics[width=8.5cm]{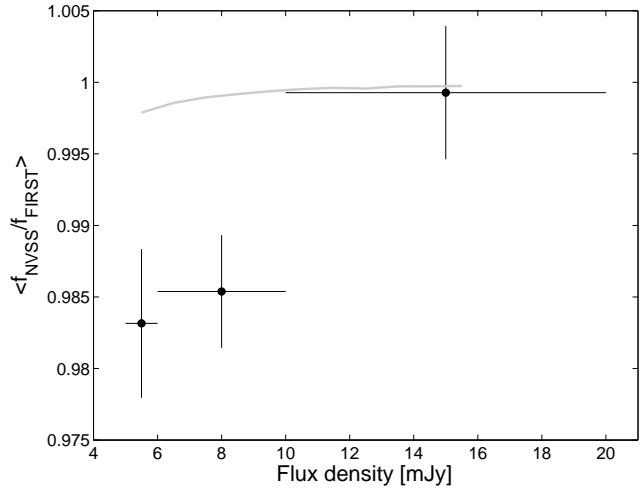}}
\caption{The mean of the flux ratio between matched individual NVSS and FIRST point sources,
as a function of flux density (black circles).
The horizontal ``error bars'' represent the bin size in which the mean
flux ratio was measured.
The expected amplitude of the Eddington-like bias (see text) is shown in gray line.
\label{Fig:FIRST_NVSS_Flux_Bias}}
\end{figure}

Next, we would like to compare the fluxes of unresolved NVSS and FIRST
sources.
However, systematic biases in the NVSS and/or FIRST flux calibration could effect
our analysis. Condon et al. (1998) and Becker et al. (1995)
discussed photometric errors such as
the CLEAN bias, and they made corrections to their source catalogs flux
densities. For both the NVSS and FIRST catalogs the
CLEAN bias is $\approx -0.3$ mJy, i.e. of order the rms noise in
the images.  There is also a well-known discrepancy between integrated fluxes
of extended sources in FIRST and NVSS, owing to resolution effects
(Blake \& Wall 2002).
In Figure~\ref{Fig:FIRST_NVSS_Flux_Bias} we show
the mean of the peak-flux ratio between matched
individual NVSS and FIRST point sources,
as a function of flux density (black circles).

This figure suggests that at the faint end, NVSS
fluxes are systematically lower
than FIRST fluxes.
Blake \& Wall (2002) already reported this effect,
although with an opposite direction and higher amplitude.
However, Blake \& Wall (2002) looked at both resolved and unresolved
sources\footnote{Dominated by resolved sources.},
while we are interested only in point sources.
Probably the most important reason for this trend is related to the fact
that the NVSS and FIRST surveys have different resolutions.
Another secondary effect is a bias similar to the Eddington bias (Eddington 1913).
This is because sources which flux expectancy value
is smaller than 5\,mJy (our FIRST flux cut)
and are detected in the FIRST survey above 5\,mJy (due to measurement errors)
have $>50$\% chance to have an NVSS flux density below 5\,mJy.
This effect is amplified by the fact that faint sources are more common,
per unit flux density, than bright sources.
The estimated amplitude of this bias, based on simulations,
is shown by the gray line in Figure~\ref{Fig:FIRST_NVSS_Flux_Bias}.

Given these results we correct the NVSS fluxes of the matched point
sources\footnote{The fluxes in Tables~\ref{Tab:FIRSTtime} and \ref{Tab:NVSStime} are not corrected for this bias.}
by the amount interpolated from the black circles in Figure~\ref{Fig:FIRST_NVSS_Flux_Bias}.
Above 20\,mJy we assume that the correction factor is 1.
We neglect the effect of the Eddington bias,
since its expected amplitude is negligible
(see Fig.~\ref{Fig:FIRST_NVSS_Flux_Bias}).
We note that the maximum amplitude of the applied correction is
smaller than the rms noise in the NVSS measurements and comparable
to rms noise of FIRST sources.
A caveat in our bias analysis 
is that this bias may also depend on the actual angular
extent of the unresolved sources.
Therefore, the bias expectation value for a given flux
may depends on a ``hidden'' parameter which value is not measured,
and cannot be entirely removed from the data.

\begin{figure}
\centerline{\includegraphics[width=8.5cm]{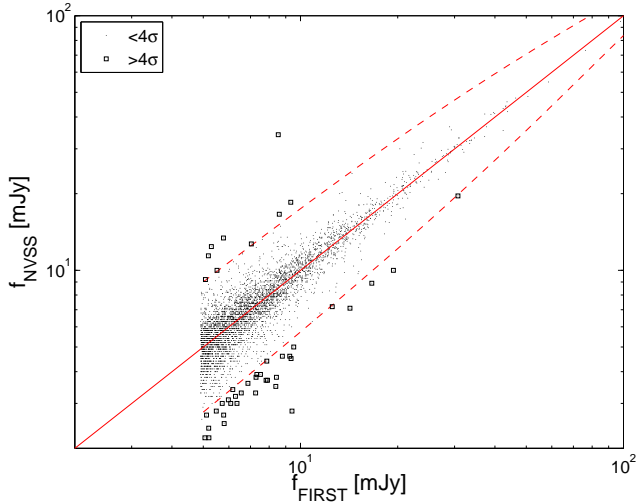}}
\caption{The FIRST peak specific flux vs. the same quantity from the NVSS
catalog
for the 4367 matched unresolved sources (see text).
The solid line shows the 1:1 line,
while the dashed lines
represent $4\sigma$ below and above
the 1:1 line.
Open boxes shows the 43 sources that are variable by more than
$4\sigma$, while all the other matched sources are marked by dots.
We note that there are more variable sources below the lower line than
above the upper line (see text).
\label{Fig:FIRST_NVSS_Flux}}
\end{figure}

The FIRST vs. NVSS peak flux-density measurements of the matched point sources
are presented in Figure~\ref{Fig:FIRST_NVSS_Flux}.
Based on this plot we estimate the standard deviation, $\sigma$, of the
differences between FIRST and NVSS
specific fluxes as a function of FIRST
flux.
This is done by calculating the 68-percentile range
in the flux-flux plot as a function of the FIRST flux density, $f_{{\rm FIRST}}$.
We divide the 68 percentile by $2$ to estimate the standard deviation,
and then fit a first order polynomial to the logarithm
of the standard deviation estimator as a function of flux density.
We find that the relative errors associated with these
difference measurements are well represented by
\begin{equation}
\sigma/f \approx 10^{-0.6652-0.0064 f_{{\rm FIRST}}}.
\label{Eq:StD}
\end{equation}
Here we define variables as objects for which the FIRST vs. NVSS flux
difference is larger\footnote{Assuming Gaussian noise,
  $4\sigma$ corresponds to
  probability of $\cong 1/15,000$ while the number of measurements in
  our experiment (number of epochs multiplied by the number of
  sources) is $8734$.} than $4\sigma$.
We found 43 such variable sources, which are listed in Table~\ref{Tab:Table_Var}.
We inspected the radio images of all these variable sources 
by eye, and comments on individual sources appear in this table.
We note that the theoretical errors are smaller, and have different functional forms,
than those implied by Equation~\ref{Eq:StD}. However, in order to avoid any possible uncertainties
in the comparison between the two catalogs, we used the
empirical errors.

Although we attempt to correct for the flux bias between
FIRST and NVSS sources,
in Figure~\ref{Fig:FIRST_NVSS_Flux}
there are more variable sources below the lower 4-$\sigma$ line than
above the upper 4-$\sigma$ line.
This systematic difference may be related to the complexity
of the bias between the FIRST and NVSS measurements, mentioned earlier.
Effectively, this systematic bias induces errors in the number of ``sigmas'' in which a
source is variable.
However,
the ratio of number of sources above the upper 4-$\sigma$ line
to that below the lower 4-$\sigma$ line,
is consistent with an additional systematic shift in the flux ratio
of $\approx 0.35\sigma$.
Therefore, we conclude that since we used a relatively large variability threshold of
4$\sigma$, most of our variable sources are probably real.

\section{The structure function}
\label{SF}

Next, we calculate the mean structure function for all the 43 sources which
exceed the 4-$\sigma$ variability threshold in
Figure~\ref{Fig:FIRST_NVSS_Flux}. 
As a reference we also calculated the structure function
for all the 3906 ``non-variable'' sources defined here as sources
which variability is less than $2\sigma$.
%
\begin{deluxetable*}{rrrrrllllllllll}
\tablecolumns{15}
\tabletypesize{\scriptsize}
\tablewidth{0pt}
\tablecaption{Variable sources}
\tablehead{
\multicolumn{2}{c}{J2000.0} &
\multicolumn{3}{c}{}            &
\multicolumn{2}{c}{USNO-B1}     &
\multicolumn{3}{c}{2MASS}       &
\multicolumn{1}{c}{{\it ROSAT}} &
\multicolumn{4}{c}{SDSS} \\
\colhead{RA}   &
\colhead{Dec}  &
\colhead{$f_{{\rm FIRST}}$}  &
\colhead{$f_{{\rm NVSS}}$}  &
\colhead{$\Delta{t}$}  &
\colhead{$B_{2}$} &
\colhead{$R_{2}$} &
\colhead{$J$}     &
\colhead{$H$}     &
\colhead{$K$}     &
\colhead{Dist}    &
\colhead{$g$}     &
\colhead{$r$}     &
\colhead{$z_{{\rm sp}}$}     &
\colhead{$z_{{\rm ph}}$}   \\
\colhead{deg}   &
\colhead{deg}  &
\colhead{mJy}  &
\colhead{mJy}  &
\colhead{day}  &
\colhead{mag} &
\colhead{mag} &
\colhead{mag}     &
\colhead{mag}     &
\colhead{mag}     &
\colhead{$''$}    &
\colhead{mag}     &
\colhead{mag}     &
\colhead{}        &
\colhead{} 
}
\startdata
 229.95525 &$-5.90828$&$19.40\pm0.14$&$10.0\pm0.5$&$ 1736.7$& \nodata&\nodata& \nodata& \nodata& \nodata& \nodata&\nodata&\nodata&  \nodata& \nodata \\
  34.81875 &$-4.64270$&$ 5.72\pm0.14$&$ 3.0\pm0.7$&$ 1270.9$& \nodata&\nodata& \nodata& \nodata& \nodata& \nodata&\nodata&\nodata&  \nodata& \nodata \\
 208.61233 &$ 0.73462$&$ 5.20\pm0.18$&$ 2.4\pm0.5$&$  760.9$& \nodata&\nodata& \nodata& \nodata& \nodata& \nodata&\nodata&\nodata&  \nodata& \nodata \\
 120.64970 &$ 7.55780$&$ 5.19\pm0.15$&$11.4\pm1.5$&$ 1758.9$& \nodata&\nodata& \nodata& \nodata& \nodata& \nodata&\nodata&\nodata&  \nodata& \nodata \\
 193.96310 &$ 8.74396$&$ 7.90\pm0.19$&$ 3.7\pm0.4$&$ 1288.4$& \nodata&\nodata& \nodata& \nodata& \nodata& \nodata&  24.70&  22.76&  \nodata&    0.30 \\
 158.12037 &$ 8.98541$&$ 8.60\pm0.15$&$16.6\pm0.6$&$ 1719.3$&   19.6 &  18.5 & \nodata& \nodata& \nodata& \nodata&  19.28&  19.28&    0.454&    0.45 \\
 155.94935 &$10.65521$&$ 5.77\pm0.15$&$13.4\pm1.6$&$ 1698.6$& \nodata&\nodata& \nodata& \nodata& \nodata& \nodata&  24.42&  22.33&  \nodata& \nodata \\
 225.14947 &$13.45289$&$ 6.08\pm0.14$&$ 3.0\pm0.4$&$ 1271.5$& \nodata&  20.0 & \nodata& \nodata& \nodata& \nodata&  23.01&  21.37&  \nodata& \nodata \\
 114.80608 &$18.03965$&$ 7.26\pm0.16$&$ 3.3\pm0.5$&$ 1762.5$&   17.1 &  16.9 &   16.14&   15.49&   14.86& \nodata&  17.36&  17.12&  \nodata&    1.00 \\
 201.32666 &$19.46406$&$ 7.04\pm0.14$&$12.7\pm0.5$&$ 1712.8$& \nodata&\nodata& \nodata& \nodata& \nodata& \nodata&\nodata&\nodata&  \nodata& \nodata \\
 206.50009 &$19.73055$&$ 5.78\pm0.15$&$ 2.7\pm0.4$&$ 1702.0$& \nodata&\nodata& \nodata& \nodata& \nodata& \nodata&  24.12&  23.18&  \nodata& \nodata \\
 244.32375 &$20.57935$&$ 7.82\pm0.14$&$ 3.7\pm0.4$&$ 1320.1$& \nodata&\nodata& \nodata& \nodata& \nodata& \nodata&  22.70&  22.29&  \nodata& \nodata \\
 222.17540 &$20.67543$&$ 6.88\pm0.15$&$ 3.6\pm0.5$&$ 1322.1$& \nodata&\nodata& \nodata& \nodata& \nodata& \nodata&\nodata&\nodata&  \nodata& \nodata \\
 247.57423 &$23.33178$&$ 9.41\pm0.36$&$ 2.8\pm0.4$&$  286.9$&   20.9 &  19.7 & \nodata& \nodata& \nodata& \nodata&  20.87&  20.52&  \nodata&    2.60 \\
 212.99535 &$23.89795$&$ 9.53\pm0.14$&$ 5.0\pm0.4$&$  284.4$& \nodata&\nodata& \nodata& \nodata& \nodata& \nodata&  22.19&  21.96&  \nodata& \nodata \\
 252.75117 &$24.26111$&$ 8.79\pm0.14$&$ 4.6\pm0.4$&$  197.4$& \nodata&\nodata& \nodata& \nodata& \nodata& \nodata&  22.86&  21.67&  \nodata& \nodata \\
 190.23317 &$24.33961$&$ 5.80\pm0.15$&$ 2.5\pm0.4$&$  284.2$& \nodata&\nodata& \nodata& \nodata& \nodata& \nodata&\nodata&\nodata&  \nodata& \nodata \\
 182.98029 &$25.36394$&$ 5.48\pm0.15$&$ 2.8\pm0.4$&$  213.8$&   20.2 &  20.5 & \nodata& \nodata& \nodata& \nodata&  20.83&  20.66&  \nodata&    0.30 \\
 247.48198 &$27.72523$&$ 7.28\pm0.16$&$ 3.8\pm0.5$&$  183.4$& \nodata&  19.7 & \nodata& \nodata& \nodata& \nodata&  22.40&  20.78&  \nodata& \nodata \\
 238.73764 &$29.95945$&$ 5.08\pm0.14$&$ 9.2\pm1.0$&$ -754.7$&   18.9 &  18.7 & \nodata& \nodata& \nodata& \nodata&  18.89&  18.87&    0.855&    0.85 \\
 203.08994 &$29.99894$&$16.62\pm0.59$&$ 8.9\pm1.1$&$ -745.7$& \nodata&\nodata& \nodata& \nodata& \nodata& \nodata&\nodata&\nodata&  \nodata& \nodata \\
 136.90323 &$30.25872$&$ 9.35\pm0.15$&$ 4.5\pm0.5$&$ -286.1$& \nodata&\nodata& \nodata& \nodata& \nodata& \nodata&  22.13&  21.99&  \nodata& \nodata \\
 203.03348 &$30.69115$&$ 9.26\pm1.35$&$ 4.6\pm0.5$&$ -735.7$& \nodata&\nodata& \nodata& \nodata& \nodata& \nodata&\nodata&\nodata&  \nodata& \nodata \\
 169.75685 &$36.87706$&$ 9.33\pm0.15$&$18.5\pm1.1$&$ -289.7$&   16.3 &  15.7 &   15.98&   15.25&   14.60& \nodata&  25.12&  19.25&  \nodata& \nodata \\
 170.99205 &$38.43646$&$ 6.17\pm0.15$&$ 3.4\pm0.4$&$ -260.7$& \nodata&\nodata& \nodata& \nodata& \nodata& \nodata&\nodata&\nodata&  \nodata& \nodata \\
 122.43647 &$38.50171$&$ 7.86\pm0.14$&$ 4.4\pm0.5$&$  202.8$& \nodata&\nodata& \nodata& \nodata& \nodata& \nodata&\nodata&\nodata&  \nodata& \nodata \\
 119.02080 &$39.25484$&$ 6.36\pm0.14$&$ 3.0\pm0.4$&$  218.6$&   19.2 &  17.6 & \nodata& \nodata& \nodata& \nodata&  20.45&  19.62&  \nodata& \nodata \\
 262.32334 &$44.77948$&$ 5.98\pm0.14$&$ 3.1\pm0.5$&$  700.7$& \nodata&\nodata& \nodata& \nodata& \nodata& \nodata&\nodata&\nodata&  \nodata& \nodata \\
 229.90208 &$44.97024$&$ 5.12\pm0.15$&$ 2.7\pm0.4$&$  710.8$& \nodata&\nodata& \nodata& \nodata& \nodata& \nodata&  24.15&  23.91&  \nodata&    0.10 \\
 134.99454 &$45.87719$&$30.74\pm0.15$&$19.6\pm0.7$&$ 1207.4$&   19.2 &  18.6 & \nodata& \nodata& \nodata& \nodata&  18.86&  19.05&    0.440&    0.50 \\
 220.45533 &$50.71170$&$ 5.06\pm0.16$&$ 2.2\pm0.4$&$  753.7$& \nodata&\nodata& \nodata& \nodata& \nodata& \nodata&\nodata&\nodata&  \nodata& \nodata \\
 262.65758 &$52.24396$&$ 7.32\pm0.14$&$ 3.9\pm0.4$&$  766.6$& \nodata&\nodata&   16.57&   15.86&   15.01& \nodata&\nodata&\nodata&  \nodata& \nodata \\
 211.65318 &$52.68479$&$ 8.42\pm0.22$&$ 3.8\pm0.4$&$  772.9$& \nodata&\nodata& \nodata& \nodata& \nodata& \nodata&\nodata&\nodata&  \nodata& \nodata \\
 159.71340 &$53.49093$&$ 8.53\pm0.15$&$34.1\pm1.8$&$  765.5$&   19.9 &\nodata& \nodata& \nodata& \nodata&    53.2&\nodata&\nodata&  \nodata& \nodata \\
 226.56440 &$53.92288$&$ 5.30\pm0.15$&$12.4\pm1.1$&$  770.9$&   21.6 &\nodata& \nodata& \nodata& \nodata& \nodata&  21.75&  21.57&  \nodata&    2.40 \\
 179.14175 &$54.63838$&$12.54\pm0.27$&$ 7.2\pm0.5$&$  779.5$& \nodata&\nodata& \nodata& \nodata& \nodata& \nodata&\nodata&\nodata&  \nodata& \nodata \\
 157.78099 &$56.75153$&$ 6.56\pm0.18$&$ 3.3\pm0.4$&$  780.2$&   20.6 &  18.9 & \nodata& \nodata& \nodata& \nodata&  20.84&  19.82&  \nodata& \nodata \\
 157.96070 &$56.85260$&$ 6.29\pm0.20$&$ 3.2\pm0.4$&$  780.2$& \nodata&\nodata& \nodata& \nodata& \nodata& \nodata&\nodata&\nodata&  \nodata& \nodata \\
 160.55682 &$59.35609$&$ 5.51\pm0.15$&$10.0\pm0.9$&$ 2642.4$& \nodata&\nodata& \nodata& \nodata& \nodata& \nodata&\nodata&\nodata&  \nodata& \nodata \\
 151.08074 &$59.53517$&$ 8.38\pm0.16$&$ 3.5\pm0.4$&$ 3075.2$& \nodata&\nodata& \nodata& \nodata& \nodata& \nodata&\nodata&\nodata&  \nodata& \nodata \\
 185.45517 &$61.59247$&$ 7.51\pm0.15$&$ 3.9\pm0.4$&$ 2655.3$& \nodata&\nodata& \nodata& \nodata& \nodata& \nodata&  23.10&  21.16&  \nodata& \nodata \\
 248.86649 &$62.64994$&$14.23\pm1.51$&$ 7.1\pm1.1$&$ 2665.3$& \nodata&\nodata& \nodata& \nodata& \nodata& \nodata&\nodata&\nodata&  \nodata& \nodata \\
 124.64543 &$63.90501$&$ 5.21\pm0.14$&$ 2.2\pm0.4$&$ 3192.3$& \nodata&\nodata& \nodata& \nodata& \nodata& \nodata&\nodata&\nodata&  \nodata& \nodata 
\enddata
\tablecomments{List of 43 sources which vary by more than 4$\sigma$ between the FIRST and NVSS epochs.
The table is sorted by declination.
{\it Columns description:} $f$ is the peak specific flux
and its error. The subscript indicate the catalog name.
$\Delta{t}$ is the time between the FIRST and NVSS observations.
The position of each source was cross correlated with various catalogs,
including the USNO-B1 (Monet al. 2003),
2MASS (Skrutskie et al. 2006),
ROSAT bright and faint source catalogs (Voges et al. 1999; Voges et al. 2000),
and the Sloan Digital Sky Survey (York et al. 2000).
In case counterparts are found we list their
USNO-B1 $B_{2}$ and $R_{2}$ magnitudes,
2MASS $J$, $H$ and $K$ magnitudes,
distance from ROSAT source,
and SDSS $g$ and $r$ magnitudes and redshifts.
We use search radius,
relative to the FIRST catalog position, of $60''$ for ROSAT and $2.5''$
for all the other catalogs.
$z$ is the SDSS spectroscopic redshift of the source, while $z_{{\rm ph}}$
is the photometric redshift of the source based on the SDSS colors.
We use a photometric redshift estimator for quasars
which is described in Ofek et al. (2002).
The photometric redshift is calculated only if the source
is indicated as a possible quasar in the SDSS database.
We note that non of these sources are associated with a GB6 source
(e.g., Gregory et al. 1996) or a known pulsar.\\
{\it Comments on individual sources:}\\
RA$=155.94935$\,deg, Dec$=+10.65521$\,deg : The peak flux we measure in the NVSS image is a factor of two lower
than the flux stated in the NVSS catalog, so this may be a constant source.\\
RA$=247.57423$\,deg, Dec$=+23.33178$\,deg : Near the strong source 3C\,340 -- NVSS and FIRST images are noisy. \\
RA$=203.03348$\,deg, Dec$=+30.69115$\,deg : Radio images are noisy.\\
RA$=159.71340$\,deg, Dec$=+53.49093$\,deg : This is possibly a radio SN in the outskirts of NGC 3310 (Argo et al. 2004).\\
RA$=226.56440$\,deg, Dec$=+53.92288$\,deg : NVSS image shows a double source.\\
RA$=157.78099$\,deg, Dec$=+56.75153$\,deg : Extended emission 15 arcmin from source.\\
RA$=157.96070$\,deg, Dec$=+56.85260$\,deg : Extended emission 4 arcmin from source.\\
RA$=124.64543$\,deg, Dec$=+63.90501$\,deg : FIRST flux may be influenced by sidelobes from 3C\,343 (4.5 Jy, 9\,arcmin to the NW).
}
\label{Tab:Table_Var}
\end{deluxetable*}
%

The structure function was calculated according to the following scheme.
For each pair of matched FIRST and NVSS measurements, we calculate
the time difference
between the FIRST ($t_{{\rm ep}}^{{\rm FIRST}}$) and NVSS
($t_{{\rm ep}}^{{\rm NVSS}}$) epochs:
$\Delta{t_{i}}=t_{{\rm ep}}^{{\rm FIRST}}-t_{{\rm ep}}^{{\rm NVSS}}$.
Here, $i$ is the index of the pair.
We also calculate for each pair
\begin{equation}
\Delta{f_{i}}/\bar{f}_{i} = \frac{f_{{\rm NVSS},i} - \bar{f}_{i}}{\bar{f}_{i}},
\label{Eq:DiffF}
\end{equation}
where $f_{{\rm NVSS},i}$ and $f_{{\rm FIRST},i}$
are the NVSS and FIRST specific fluxes of the $i$-th source,
and $\bar{f}_{i} = (f_{{\rm NVSS},i} + f_{{\rm FIRST},i})/2$.
Next, the structure function and its error are estimated
in bins of 500 days, between zero and 3500\,days,
by calculating the mean and standard deviation of $\Delta{f_{i}}/\bar{f}_{i}$
for all the pairs $i$ in the appropriate bin.

\begin{figure}
\centerline{\includegraphics[width=8.5cm]{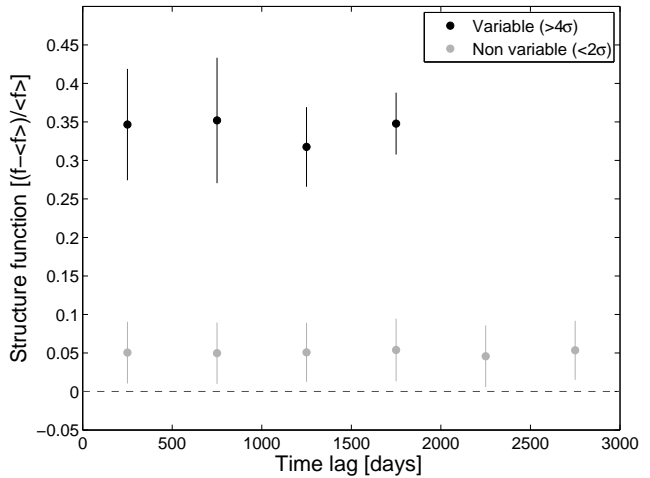}}
\caption{The structure function of variable sources (black circles)
and non-variable sources (gray circles)
as a function of time lag.
The measurements are displayed only for bins in which the number
of flux differences is larger than four.
The errors are based on the standard deviation in each bin.
The horizontal dashed line marks the zero structure function.
\label{Fig:FIRST_NVSS_SF}}
\end{figure}

Figure~\ref{Fig:FIRST_NVSS_SF} shows the structure function
for the variable (black) and non variable (gray)
sources.
The structure function in this time range is consistent with being
flat, with mean relative variability of about 35\%.
However, the value of 35\% probably represents our sensitivity for
variability rather than some physical variability level. The only 
physically interesting fact is the flatness of the structure function.

\section{Discussion}
\label{Disc}

We present versions of the FIRST and NVSS catalogs that
contain the mean epoch in which each source was observed.
We use these catalogs to look for variable sources,
and we construct the structure function
for these objects.
We show that the structure function is flat on time scales
between about half a year and five years.

It is well known that the structure function
of variable radio sources rises
on time scales of days to tens of days
(e.g., Qian et al. 1995; Gaensler \& Hunstead 2000;
Lovell et al. 2008; Ofek et al. 2011).
Intrinsic variability of compact radio sources,
which are mainly Active Galactic Nuclei (AGN),
on days time scales would imply
that the sources have small physical size.
This in turn requires a very high
rest-frame brightness temperature ($T_{{\rm B, rest}}$),
orders of magnitude above $\sim 10^{12}$\,K which is the limit for
an incoherent synchrotron source
(e.g., Kellermann \& Pauliny-Toth 1969; Readhead 1994).
Therefore, most or all of the variability of radio
sources, below $\sim5$\,GHz, on these short time scales is presumably due
to scintillations in the ISM.
Moreover, based on causality arguments, intrinsic variability of compact radio sources
is expected only on time scales ($\tau_{V}$) larger than
\begin{eqnarray}
\tau_{V} & \gtorder & 50 \Big(\frac{\Delta{f}}{1\,{\rm mJy}} \Big)^{1/2}
                         \Big(\frac{d_{{\rm lum}}}{6.7\,{\rm Gpc}} \Big)
                         \Big(\frac{\nu}{1.4\,{\rm GHz}}           \Big)^{-1} \cr
     &          & \times \Big(\frac{\mathcal{D}}{10}               \Big)^{-3/2}
                         \Big(\frac{1+z}{2}                        \Big)^{3/2}
                         \Big(\frac{T_{{\rm B,rest}}}{10^{12}\,{\rm K}} \Big)^{-1/2}\,{\rm day},
\label{TimeScaleTb}
\end{eqnarray}
where $\Delta{f}$ is the variation amplitude in specific flux,
$d_{{\rm lum}}$ is the luminosity distance (normalized at $z\cong1$),
$\mathcal{D}$ is the Doppler factor of
a relativistic motion in the source, and $z$ is the source redshift.
In Equation~\ref{TimeScaleTb},
 $T_{{\rm B,rest}}$ is given in the rest frame and all the other parameters
are in the observer reference frame.
We note that Eq.~\ref{TimeScaleTb} is 
derived from the relation
for the brightness temperature, and replacing the source size
by $c\tau_{V}$, where $c$ is the speed of light.

In contrast, observations of quasars and BL Lac objects
performed in the 4.8--14.5\,GHz range showed
that the structure function saturates
only on time scales between a year and ten
years (Hughes, Aller \& Aller 1992).
Therefore,
the fact that we do~not see any significant rise in the structure function
on time scales of months to years
suggests that, at 1.4\,GHz, the amplitude of intrinsic variability
relative to scintillations is small.
Alternatively, it may suggest that the 1.4\,GHz power spectrum of AGN
radio variability is consistent with a white-noise power spectrum,
rather than the red-noise power spectrum typically
seen at shorter wavelengths (e.g., Giveon et al. 1999; Markowitz et al. 2003).
We note that Padrielli et al. (1987) reported
on a class of sources (denoted ``C-BBV'' in their terminology)
which show correlated intrinsic variability in low (0.4\,GHz)
and high (14.5\,GHz) frequencies.
However, these sources are a minority among the variable sources in their sample.

\begin{figure}
\centerline{\includegraphics[width=8.5cm]{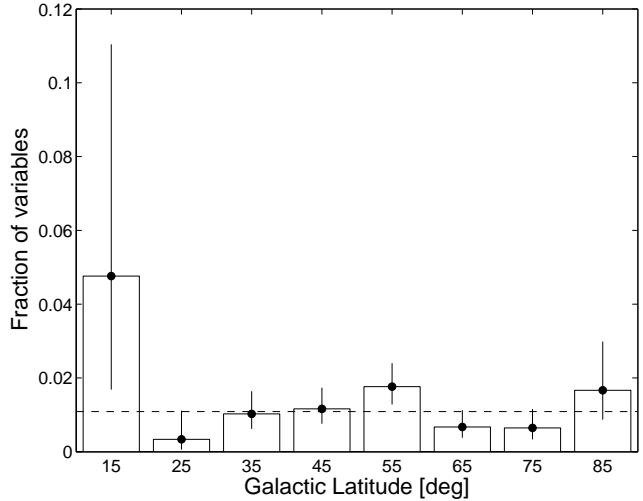}}
\caption{The fraction of
variables, relative to all the non-variable and variable sources,
as a function of Galactic latitude.
The error bars represent the 1-$\sigma$ errors
on the fractions (Gehrels 1986).
The dashed horizontal line marks the mean ``all-sky'' fraction
of variables in our sample (i.e., $43/3949=0.0109$).
\label{Fig:GalLat_FracVar}}
\end{figure}

Figure~\ref{Fig:GalLat_FracVar}
shows the fraction of variables as a function of Galactic latitude.
The first bin contains two variables out of 42 objects (a fraction of 0.048).
However, the expectation value in this bin, estimated based on the mean 
fraction of variables (dashed line in Fig.~\ref{Fig:GalLat_FracVar}; $0.0109$)
multiplied by the number 
of sources in the first bin (42) is 0.457.
Assuming a binomial distribution, the cumulative probability to observe
$\ge2$ events, given an expectation value of 0.457, is 7.7\%.
This rules out the null hypothesis that the low-latitude
variable-fraction is drawn from a uniform all-sky distribution
at the 92.3\% confidence.
Therefore, a larger sample is required in order to confirm
the earlier claims that the fraction of variables is
larger at low Galactic latitude (e.g., Gaensler \& Hunstead 2000).
If this excess is real, then a
plausible explanation is that it is due to ISM scintillations which
are more prominent at low Galactic latitudes.  However, we cannot rule
out that some of this excess in variability is due to a population of
Galactic variable sources as suggested by Becker et al. (2010).

\begin{figure}
\centerline{\includegraphics[width=8.5cm]{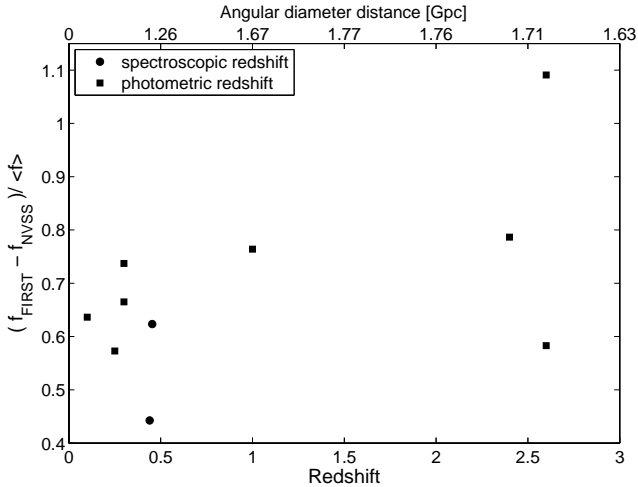}}
\caption{The variability relative amplitude, $\Delta{f_{i}}/\bar{f}_{i}$,
as a function of redshift for all nine variable sources that
are quasars or quasar candidates (Table~\ref{Tab:Table_Var}).
The upper x-axis shows the angular diameter distance corresponding to
the redshift, assuming WMAP fifth year cosmological
parameters (Komatsu et al. 2009).
\label{Fig:z_amplitude}}
\end{figure}

Finally, we use our dataset to look for correlation
of the variability amplitude with redshift.
There are several factors that can contribute to such a correlation.
For example, scintillations depends on the source
angular size (larger amplitude for smaller sources),
intrinsic source size evolution, broadening due to scattering,
and maybe even scintillation in the intra-galactic medium
(which may depends on the He re-ionization).
Figure~\ref{Fig:z_amplitude} shows $\Delta{f_{i}}/\bar{f}_{i}$
as a function of redshift
for all nine variable sources for which we have a redshift
estimate (Table~\ref{Tab:Table_Var}).
The Spearman rank (Pearson)
correlation coefficient between the redshift and the relative
variability amplitude is 0.59 (0.79).
In order to estimate the significance of this correlation
we conduct $10^{4}$ bootstrap simulations
(Efron 1982; Efron \& Tibshirani 1993).
In each simulation we select, for each source redshift,
a random variability amplitude from the list of nine amplitudes.
We find that the probability to get a Spearman rank correlation
coefficient $>0.59$ is about 5\%.
Therefore, unlike
Lovell et al. (2008) we do~not
find any strong evidence for correlation between scintillations and redshift.
However, our sample is considerably smaller than the
one presented by Lovell et al. (2008).
We note that eight out of the nine sources in
Figure~\ref{Fig:z_amplitude} are found
above Galactic latitude of 40\,deg,
and one source is at Galactic latitude of about 18\,deg.
Removing the single low galactic latitude source
(the source at $z\approx1$ in Figure~\ref{Fig:z_amplitude})
reduces the Spearman rank correlation to 0.42
and therefore does~not change this results significantly.

\acknowledgments
We thank Orly Gnat for reading the manuscript and to Jim Condon and an anonymous referee for
their comments.
EOO is supported by an Einstein fellowship and NASA grants.
This paper is based on observations conducted with the VLA,
which is
operated by the National Radio Astronomy Observatory (NRAO),
a facility of the National Science Foundation operated under
cooperative agreement by Associated Universities, Inc.


\begin{thebibliography}{}

\bibitem[Argo et al.(2004)]{2004MNRAS.351L..66A} Argo, M.~K., Muxlow, 
T.~W.~B., Pedlar, A., Beswick, R.~J., \& Strong, M.\ 2004, MNRAS, 351, L66 

\bibitem[Becker et al.(1995)]{1995ApJ...450..559B} Becker, R.~H., White, 
R.~L., \& Helfand, D.~J.\ 1995, ApJ, 450, 559 

\bibitem[Becker et al.(2010)]{2010arXiv1005.1572B} Becker, R.~H., Helfand, 
D.~J., White, R.~L., \& Proctor, D.~D.\ 2010, arXiv:1005.1572 








\bibitem[Blake 
\& Wall(2002)]{2002MNRAS.337..993B} Blake, C., \& Wall, J.\ 2002, MNRAS, 337, 993 



\bibitem[Blandford \& Narayan(1985)]{1985MNRAS.213..591B} Blandford, R., \& Narayan, R.\ 1985, MNRAS, 213, 591 

\bibitem[Blandford et al.(1986)]{1986ApJ...301L..53B} Blandford, R., 
Narayan, R., \& Romani, R.~W.\ 1986, ApJL, 301, L53 
























\bibitem[Condon et al.(1979)]{1979AJ.....84....1C} Condon, J.~J., Ledden, 
J.~E., Odell, S.~L., \& Dennison, B.\ 1979, AJ, 84, 1 



\bibitem[Condon et al.(1998)]{1998AJ....115.1693C} Condon, J.~J., Cotton, 
W.~D., Greisen, E.~W., Yin, Q.~F., Perley, R.~A., Taylor, G.~B., \& 
Broderick, J.~J.\ 1998, \aj, 115, 1693 













\bibitem[Eddington(1913)]{1913MNRAS..73..359E} Eddington, A.~S.\ 1913, MNRAS, 73, 359 


\bibitem[Efron(1982)]{Efron1982} Efron, B., 1982, The Jackknife, the Bootstrap and Other Resampling Plans, The Society for Industrial and Applied Mathematics

\bibitem[ET(1993)]{ET1993} Efron, B., Tibshirani, R.J., 1993, An introduction to the bootstrap, Monographs on statistics and applied probability 57, Chapman \& Hall













\bibitem[Gaensler \& Hunstead(2000)]{2000PASA...17...72G} Gaensler, B.~M., \& Hunstead, R.~W.\ 2000, PASA, 17, 72 



\bibitem[Gal-Yam et al.(2006)]{2006ApJ...639..331G} Gal-Yam, A., et al.\ 
2006, ApJ, 639, 331 

\bibitem[Gehrels(1986)]{1986ApJ...303..336G} Gehrels, N.\ 1986, ApJ, 303,
336



\bibitem[Ghosh \& Rao(1992)]{1992A&A...264..203G} Ghosh, T., \& Rao, A.~P.\ 1992, A\&A, 264, 203 

\bibitem[Giveon et al.(1999)]{1999MNRAS.306..637G} Giveon, U., Maoz, D., Kaspi, S., Netzer, H., \& Smith, P.~S.\ 1999, MNRAS, 306, 637




\bibitem[Goodman \& Narayan(1985)]{1985MNRAS.214..519G} Goodman, J., \& Narayan, R.\ 1985, MNRAS, 214, 519 

\bibitem[Goodman et al.(1987)]{1987MNRAS.229...73G} Goodman, J.~J., Romani, 
R.~W., Blandford, R.~D., \& Narayan, R.\ 1987, MNRAS, 229, 73 








\bibitem[Gregory et al.(1996)]{1996ApJS..103..427G} Gregory, P.~C., Scott, 
W.~K., Douglas, K., \& Condon, J.~J.\ 1996, ApJS, 103, 427 












\bibitem[Helfand et al.(1996)]{1996ASPC..110..214H} Helfand, D.~J., Das, 
S.~R., Becker, R.~H., White, R.~L., \& McMahon, R.~G.\ 1996, Blazar Continuum Variability, 110, 214 



\bibitem[Hjellming \& Narayan(1986)]{1986ApJ...310..768H} Hjellming, R.~M., \& Narayan, R.\ 1986, ApJ, 310, 768 

\bibitem[Hughes et al.(1992)]{1992ApJ...396..469H} Hughes, P.~A., Aller, 
H.~D., \& Aller, M.~F.\ 1992, ApJ, 396, 469 

\bibitem[Hunstead(1972)]{1972ApL....12..193H} Hunstead, R.~W.\ 1972, 
ApL, 12, 193 










\bibitem[Kellermann 
\& Pauliny-Toth(1969)]{1969ApJ...155L..71K} Kellermann, K.~I., \& Pauliny-Toth, I.~I.~K.\ 1969, ApJL, 155, L71 



\bibitem[Komatsu et al.(2009)]{2009ApJS..180..330K} Komatsu, E., et al.\ 
2009, ApJS, 180, 330 

\bibitem[Lazio et al.(2008)]{2008ApJ...672..115L} Lazio, T.~J.~W., Ojha, 
R., Fey, A.~L., Kedziora-Chudczer, L., Cordes, J.~M., Jauncey, D.~L., 
\& Lovell, J.~E.~J.\ 2008, \apj, 672, 115 


\bibitem[Levinson et al.(2002)]{2002ApJ...576..923L} Levinson, A., Ofek, 
E.~O., Waxman, E., \& Gal-Yam, A.\ 2002, ApJ, 576, 923 




\bibitem[Lovell et al.(2008)]{2008ApJ...689..108L} Lovell, J.~E.~J., et 
al.\ 2008, \apj, 689, 108 



\bibitem[Markowitz et al.(2003)]{2003ApJ...593...96M} Markowitz, A., et 
al.\ 2003, ApJ, 593, 96 



\bibitem[McLaughlin et al.(2006)]{2006Natur.439..817M} McLaughlin, M.~A., 
et al.\ 2006, Nature, 439, 817 


\bibitem[Mitchell et al.(1994)]{1994ApJS...93..441M} Mitchell, K.~J., 
Dennison, B., Condon, J.~J., Altschuler, D.~R., Payne, H.~E., O'dell, 
S.~L., \& Broderick, J.~J.\ 1994, ApJS, 93, 441 

\bibitem[Monet et al.(2003)]{2003AJ....125..984M} Monet, D.~G., et al.\ 
2003, AJ, 125, 984 


\bibitem[Ofek et al.(2002)]{2002MNRAS.337.1163O} Ofek, E.~O., Rix, H.-W., 
Maoz, D., \& Prada, F.\ 2002, MNRAS, 337, 1163 





\bibitem[Ofek et al.(2010)]{2010ApJ...711..517O} Ofek, E.~O., Breslauer, 
B., Gal-Yam, A., Frail, D., Kasliwal, M.~M., Kulkarni, S.~R., 
\& Waxman, E.\ 2010, ApJ, 711, 517 

\bibitem[Ofek et al.(2011)]{2011Ofek} Ofek, E.~O., et al., submitted to ApJ

\bibitem[Padrielli et 
al.(1987)]{1987A&AS...67...63P} Padrielli, L., et al.\ 1987, A\&AS, 67, 63 

\bibitem[Qian et al.(1995)]{1995A&A...295...47Q} Qian, S.~J., Britzen, S., Witzel, A., Krichbaum, T.~P., Wegner, R., \& Waltman, E.\ 1995, A\&A, 295, 47 

\bibitem[Readhead(1994)]{1994ApJ...426...51R} Readhead, A.~C.~S.\ 1994, ApJ, 426, 51 

\bibitem[Rickett(1990)]{1990ARA&A..28..561R} Rickett, B.~J.\ 1990, ARA\&A, 28, 561 

\bibitem[Rickett et al.(1984)]{1984A&A...134..390R} Rickett, B.~J., Coles, W.~A., \& Bourgois, G.\ 1984, A\&A, 134, 390 

\bibitem[Rys \& Machalski(1990)]{1990A&A...236...15R} Rys, S., \& Machalski, J.\ 1990, A\&A, 236, 15 

\bibitem[Skrutskie et al.(2006)]{2006AJ....131.1163S} Skrutskie, M.~F., et 
al.\ 2006, AJ, 131, 1163 

\bibitem[Spangler et 
al.(1989)]{1989A&A...209..315S} Spangler, S., Fanti, R., Gregorini, L., \& Padrielli, L.\ 1989, A\&A, 209, 315 

\bibitem[Taylor \& Gregory(1983)]{1983AJ.....88.1784T} Taylor, A.~R., \& Gregory, P.~C.\ 1983, AJ, 88, 1784 

\bibitem[Voges et al.(1999)]{1999A&A...349..389V} Voges, W., et al.\ 1999, A\&A, 349, 389 

\bibitem[Voges et al.(2000)]{2000IAUC.7432....3V} Voges, W., et al.\ 2000, IAUC, 7432, 3 

\bibitem[York et al.(2000)]{2000AJ....120.1579Y} York, D.~G., et al.\ 2000, 
AJ, 120, 1579 


\end{thebibliography}
\end{document}